\documentclass[10pt,a4paper]{article}
\usepackage{amsmath}
\usepackage{amssymb}
\usepackage{graphicx}

\usepackage{natbib}

\bibpunct{(}{)}{;}{a}{}{,}  

\makeatletter
\newdimen\bibindent
\def\@biblabel#1{}
\def\@cite#1#2{{#1\if@tempswa , #2\fi}}

\renewenvironment{thebibliography}[1]
     {
        \small 
      \list{\@biblabel{\@arabic\c@enumiv}}%
           {\settowidth\labelwidth{\@biblabel{#1}}%
           \leftmargin\bibindent
           \setlength{\itemindent}{-\leftmargin}
           \setlength{\parsep}{0pt}
           \usecounter{enumiv}%
           \renewcommand\theenumiv{\@arabic\c@enumiv}
}
      \sloppy\clubpenalty4000\widowpenalty4000%
      \sfcode`\.\@m}
     {\def\@noitemerr
       {\@latex@warning{Empty `thebibliography' environment}}%
      \endlist}

\makeatother
\begin{document}

\setcounter{figure}{0}
\setcounter{table}{0}
\setcounter{footnote}{0}
\setcounter{equation}{0}

\vspace*{0.5cm}

\noindent {\Large TESTS OF GRAVITATION AT SOLAR SYSTEM SCALES BEYOND THE PPN FORMALISM}
\vspace*{0.7cm}

\noindent\hspace*{1.5cm} A. HEES$^1$, W. M. FOLKNER$^1$, R. A. JACOBSON$^1$, R. S. PARK$^1$, B. LAMINE$^2$, \\C. LE PONCIN-LAFITTE$^3$, P. WOLF$^3$\\
\noindent\hspace*{1.5cm} $^1$ Jet propulsion Laboratory - California Institute of technology\\
\noindent\hspace*{1.5cm} 4800 Oak Grove Drive, Pasadena, CA, 91109, USA\\
\noindent\hspace*{1.5cm} e-mail: aurelien.hees@gmail.com \\
\noindent\hspace*{1.5cm} $^2$ IRAP, Universit\'e de Toulouse, CNRS, 31400 Toulouse, France\\
\noindent\hspace*{1.5cm} $^3$  Observatoire de Paris, SYRTE, CNRS/UMR 8630, UPMC\\
\noindent\hspace*{1.5cm}  61 avenue de l'Observatoire, F-75014 Paris, France\\

\vspace*{0.5cm}

\noindent {\large ABSTRACT.} In this communication, the current tests of gravitation available at Solar System scales are recalled. These tests rely mainly on two frameworks: the PPN framework and the search for a fifth force. Some motivations are given to look for deviations from General Relativity in other frameworks than the two extensively considered. A recent analysis of Cassini data in a MOND framework is presented. Furthermore, possibilities to constrain Standard Model Extension parameters using Solar System data are developed.

\vspace*{1cm}

\noindent {\large 1. INTRODUCTION}

\smallskip
The General theory of Relativity (GR) is not the ultimate theory of gravity. This assumption is motivated by theoretical developments like the search for a quantum theory of gravitation or the unification of the gravitation with the other fundamental interactions. From an observational point of view, puzzling galactic and cosmological observations can not be explained by GR and the standard model of particles. Accounting for these observations is usually done by introducing new types of matter: the so-called Dark Matter and Dark Energy fluids. Since these constituents have not been directly observed yet, they may be a hint of a deviation from GR. Therefore, searches for alternative theories of gravity is very important. Due to the very high accuracy measurements available, the Solar System is a very good laboratory to test GR. In Section 2, we will briefly recall what are the two formalisms widely used so far to test GR in the Solar System and what are the current constraints in these formalisms. In Section 3, we give some motivations to consider tests of gravitation beyond the standard formalisms. We also present a sensitivity analysis of Messenger and Cassini data to the Standard Model Extension. Finally, in Section 4, we present a new test of the MOND phenomenology done using Cassini data.

\vspace*{0.7cm}
\noindent {\large 2. STANDARD TESTS OF GENERAL RELATIVITY}
\smallskip

GR is built on two principles. The first one is called the Einstein Equivalence Principle. It implies that gravity can be identified to space-time geometry which is mathematically described by a space-time metric $g_{\mu\nu}$. In particular, test masses follow geodesics of this metric and ideal clocks measure the proper time of this metric. This principle is now very well tested through the Universality of Free Fall, tests of the Local Lorentz Invariance and tests of the Local Position Invariance. A review of the current tests of the Einstein Equivalence Principle can be found in \citet{will:2006cq}. Nevertheless, there exist strong theoretical motivations to improve the current constraints on the equivalence principle \citep{damour:2012zr} coming from string theories, from theories with variable fundamental constants, from the anthropic principle, \dots

The second principle upon which GR is built is the Einstein field equations that determine the form of the metric tensor. In GR, the form of the metric tensor is directly influenced by the energy/matter content of the space-time through the Einstein equations. Up to now, two formalisms have been widely used to test the form of the metric tensor: the Parametrized Post-Newtonian (PPN) formalism and the search for a fifth force.

The PPN formalism is a phenomenology making an interface between theoretical developments and experiments. In the PPN formalism (fully described in \citet{will:1993fk}), the space-time metric is parametrized by 10 dimensionless coefficients that can be tested independently of any fundamental underlying theory. Forty years of precise experiments have constrained PPN parameters to be very close to those expected from GR (for a review, see \citet{will:2006cq}). In particular, the two most famous PPN parameters $\gamma$ and $\beta$ (whose values are 1 in GR) are now constrained at the level of $10^{-5}$ with the measure of the Shapiro delay with the Cassini spacecraft \citep{bertotti:2003uq} and with planetary ephemerides \citep{pitjeva:2013fk,verma:2014jk}.

The fifth force formalism considers a modification of the Newtonian potential with a Yukawa potential parametrized by a range of interaction $\lambda$ and a strength of interaction $\alpha$ (see \citet{talmadge:1988uq,adelberger:2009fk}).   These parameters have been tested in a very wide range (see \citet{adelberger:2009fk} and Fig. 31 of \citet{konopliv:2011dq}). In particular, $\alpha$ is constrained at a very high level of accuracy ($\alpha<10^{-10}$) at Earth-Moon and Sun-Mars distances. It is also worth mentioning that windows remain opened at very short distances and at very large distances.

While the current constraints on these two formalisms are very impressive, there exist theoretical models that predict deviations smaller than the current accuracy. One good example of such a model is given by theories developing screening mechanism (see for example \citet{brax:2012yg,hees:2012kx}). Upcoming space missions like Gaia or BepiColombo will improve the current limits in these formalisms.

\vspace*{0.7cm}
\noindent {\large 3. BEYOND THE STANDARD FORMALISMS}
\smallskip

Even if the PPN and the fifth force formalisms have been extensively used so far, there are motivations to consider alternative frameworks. Indeed, not all the alternative theories of gravity enter these formalisms. To illustrate this, we will give four examples of alternative theories of gravity considered in the literature that are not entering the standard formalisms. 

First, the \emph{Post-Einsteinian Gravity (PEG)} \citep{jaekel:2005zr,jaekel:2006kx} is an alternative theory of gravity based on a non local extension of Einstein field equations as suggested by radiative corrections. Phenomenologically in the Solar System, the space-time metric can be parametrized by two radial dependent potentials. This can be seen as an extension of the PPN formalism where the $\gamma$ and $\beta$ PPN parameters have been promoted to a function of the radial coordinate.

The second example is given by the \emph{Standard Model Extension (SME)} framework. This phenomenology has been developed to systematically consider possible violations of the Lorentz symmetry. In the gravitational sector, the weak field space-time metric is parametrized by a symmetric trace-free tensor $\bar s_{\mu\nu}$ \citep{bailey:2006uq} different from the PPN parametrization.

Another example is given by the \emph{Fab Four theory}, which is a tensor-scalar theory of gravitation developed in the cosmological context to solve the cosmological constant problem. It was shown that the Solar System metric derived from this theory is parametrized by four parameters and is not covered by the PPN parametrization \citep{bruneton:2012kx}.

Finally, it was shown that the \emph{MOdified Newtonian Dynamic (MOND)} phenomenology produced an effect in the Solar System called the External Field Effect (EFE).  It is due to the non-linearity of MOND equations in which the gravitational dynamics of a system are influenced by the external gravitational field \citep{blanchet:2011ys}. In the Solar System, it implies a quadrupolar correction to the Newtonian potential
\begin{equation}\label{MOND}
	\Phi=-\frac{GM}{r}-\frac{Q_2}{2}x^ix^j\left(e_ie_j-\frac{1}{3}\delta_{ij}\right)
\end{equation}
where $e_i$ is a unitary vector pointing towards the galactic center and $Q_2$ is a parameter related to the fundamental MOND theory. 

The four examples given here are interesting alternative theories of gravity that do not enter the PPN or the fifth force formalism. For these kind of theories, only very few constraints at Solar System scales are available today and there is a strong interest to constrain these theories with Solar System data. 

Recently, software aiming at simulating range, Doppler and astrometric observables directly from the space-time metric has been developed \citep{hees:2012fk}. The main advantage of this software is that it allows one to produce simulations in a wide class of alternative theories of gravitation very easily. In particular, all the potential effects produced by a deviation from GR are automatically taken into account including the effect on the orbits of bodies,  on clock behavior, effect on light propagation \citep{hees:2014fk}, \dots This software also provides a fit of the initial conditions of the different bodies to identify the incompressible signature produced by a modification of gravity on observations. By incompressible signature, we refer to the signature that is not correlated with the initial conditions and that will not be absorbed in real data analysis. This signature is the one that would be observed in residuals of real data analysis (done in GR using standard procedure) while the correct gravitation theory is the considered alternative theory.

For example, we used the software described above to identify the incompressible signatures produced by SME on Messenger (around Mercury) and Cassini (around Saturn) spacecraft. The radioscience observations of these spacecraft depend on linear combinations of the fundamental parameters $\bar s_{\mu\nu}$. Figures representing the incompressible signatures produced by linear combinations of the SME parameters on Messenger and Cassini observations are given in \citet{hees:2013kx}. These signatures are characteristic of SME and can be searched in the residuals of real data analysis. 

A comparison of the amplitude of these signatures with the accuracy of the measurements gives an estimate of the uncertainties on the SME coefficients that would be reachable in a real data analysis. Under the assumptions that no anomalous residuals have been observed during the data analysis, these uncertainties can be interpreted as an order of magnitude of an upper bound on the SME coefficients. The estimated uncertainties on SME coefficients reachable using Messenger and Cassini are given in Table~\ref{tab}. These values are very good compared to the current limit on SME parameters and this gives a strong motivation to consider a test of SME using Messenger and/or Cassini radio tracking data.
\begin{table}[h]
\begin{center}
\begin{tabular}{cc}
\hline
Coeff. & Uncertainties\\\hline
$\bar s_A$ & $1.1 \times 10^{-10}$\\
$\bar s_{TX}$ & $3.1 \times 10^{-8\phantom{0}}$ \\ 
$\bar s_B$ & $1.4 \times 10^{-8\phantom{0}}$ \\
$\bar s_C$ & $3.2 \times 10^{-11}$ \\\hline
\multicolumn{2}{c}{(a) : Messenger } \\
\end{tabular}
\hspace*{3cm}
\begin{tabular}{cc}
\hline
Coeff. & Uncertainties \\\hline
$\bar s_F$ & $8.6 \times 10^{-11}$\\
$\bar s_{TX}$ & $1.2 \times 10^{-8\phantom{0}}$ \\ 
$\bar s_G$ & $1.5 \times 10^{-8\phantom{0}}$ \\
$\bar s_H$ & $2.3 \times 10^{-11}$ \\\hline 
\multicolumn{2}{c}{(b) : Cassini} \\
\end{tabular}
\end{center}
\caption{Estimated reachable uncertainties on SME coefficients. The expression of the linear combinations are given in \citet{hees:2013kx}.}
\label{tab}
\end{table}

\vspace*{0.7cm}
\noindent {\large 4. A TEST OF MOND WITH CASSINI DATA}
\smallskip

As mentioned above, the main effect of the MOND phenomenology in the Solar System is called the External Field Effect (EFE) and is modeled by a quadrupolar correction to the Newtonian potential (\ref{MOND}). This effect is parametrized by the MOND EFE parameter $Q_2$ that depends on the fundamental MOND theory (more precisely on the MOND interpolating function). The value of $Q_2$ for standard MOND interpolating functions has been computed by \citet{blanchet:2011ys} and is framed by
\begin{equation}\label{Q2t}
		2.1 \times 10^{-27} \ s^{-2} \leq Q_2 \leq 4.1 \times 10^{-26} \ s^{-2} .
\end{equation}

In a recent study \citep{hees:2014jk}, we have used 9 years of Cassini radio tracking to constrain $Q_2$. The analysis is done in two steps: first, the spacecraft trajectory around the Saturn barycenter is fitted and then the Saturn orbit is adjusted. The model used for the spacecraft data reduction is standard and can be found in \citet{moyer:2000uq}. With current models, the range and Doppler can be fit to their intrinsic noise level without any signature remaining (when using both range and Doppler). This is due to the number of free parameters that are estimated (in particular the numerous maneuvers). For this study, we have estimated spacecraft trajectories with only Doppler and satellite imaging data. Omitting the range data from the spacecraft trajectory estimates leads to larger range residuals since they are not absorbed in the spacecraft orbit parameters and this allows the range data to be used to estimate corrections to the Saturnian orbit.

The range measurements have been used to estimate corrections to the orbit of Saturn and the EFE parameter. A standard model for the adjustment has been considered and extended to take into account the EFE \citep{hees:2014jk}. The principal estimated parameters included in the fit are the orbits of the Earth and Saturn, the mass parameter of the Sun, the EFE parameter $Q_2$, a constant correction to the radio delay, a constant scale factor correction to the solar plasma modelling.

\begin{figure}[h]
\begin{center}
\includegraphics[width=0.355\linewidth]{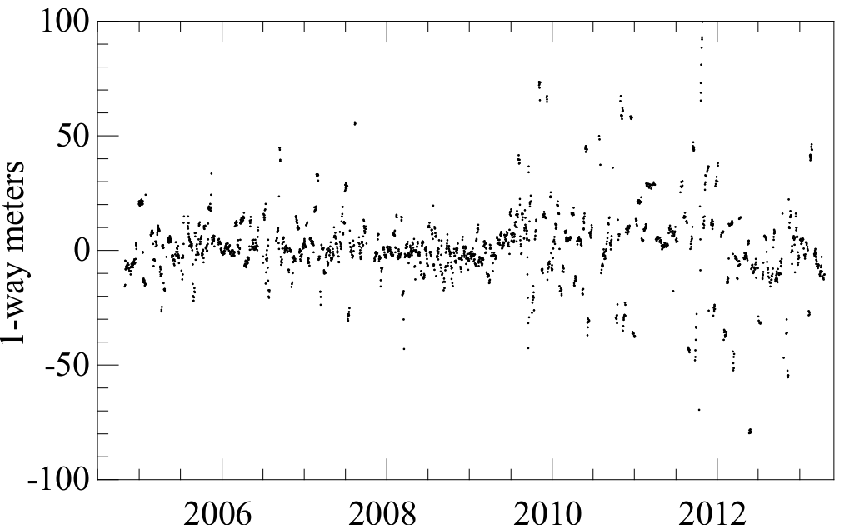}\hspace*{2cm}
\includegraphics[width=0.35\linewidth]{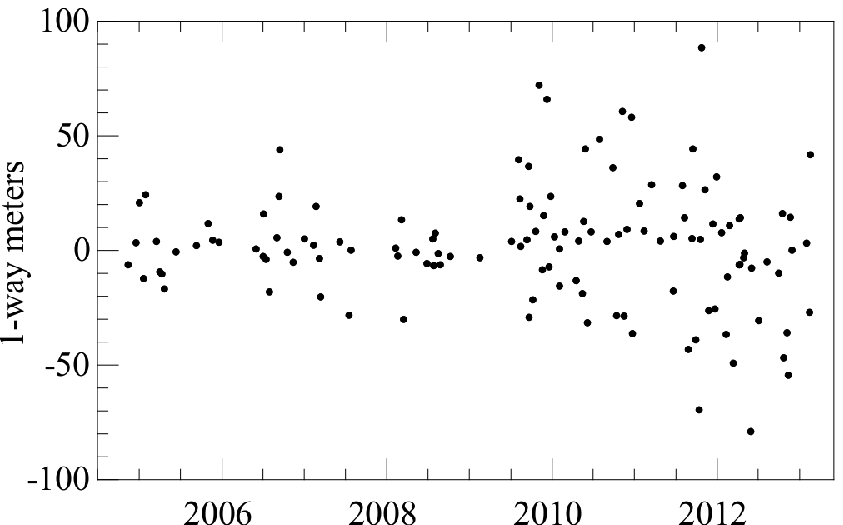}
\end{center}
\caption{Range measurement residuals (left: one observation per tracking pass - right: one observation per orbit segment).}
\label{fig}
\end{figure}

In order to have realistic uncertainties on the estimated parameters, it is important to consider observations that are independent in the adjustment. The range residuals when taking one range observations by tracking pass for the fit is presented on the left of Figure~\ref{fig}. As one can see, the residuals are clustered in groups of data corresponding to the same spacecraft orbit estimate. This means range measurements within a single spacecraft orbit segment are highly correlated since they share a common estimated spacecraft trajectory relative to Saturn. Therefore, we consider only one range observation per spacecraft segment and the corresponding residuals are presented on the right of Figure~\ref{fig}.

We obtain an estimate of the $Q_2$ parameter given by
\begin{equation}
	Q_2=(3 \pm 3)\times 10^{-27} \ s^{-2}.
\end{equation}
The value $Q_2=0$ is included in the 1-$\sigma$ confidence interval. This means the set of data used does not favor a MOND theory with respect to GR. Moreover, our result puts a very stringent constraint on the interval~(\ref{Q2t}) computed theoretically. In particular, MOND theories characterized by standard MOND interpolating functions like $\mu_{1,2}$, $\mu_{\rm exp}$ or $\mu_{\rm TeVeS}$ (see \citet{blanchet:2011ys} for a review of these MOND interpolating functions) are excluded by Cassini data.

\vspace*{0.7cm}
\textit{Acknowledgements.} The research described in this poster was partially carried out at the Jet Propulsion Laboratory, California Institute of Technology, under contract with the National Aeronautics and Space Administration \copyright \ 2013. All rights reserved. A. Hees thanks the BAEF for financial support.
\vspace*{0.7cm}

\noindent {\large 5. REFERENCES}

{

\leftskip=5mm
\parindent=-5mm

\smallskip

\bibliographystyle{aa}  
\bibliography{../../../../../Dropbox/JPL/JPL_byMe/biblio} 




















}

\end{document}